\documentclass[12pt]{article}

\usepackage[cp866]{inputenc}
\usepackage{epsfig}

\usepackage{epsfig}

\begin{document}

\setcounter{page}{1}

\vspace{.5cm}

\begin{center}
{\LARGE \bf 
On the Structure of the Magnetic Field \\
Near a Black Hole in Active Galactic Nuclei}
\end{center}

\begin{center}
V.S.~Beskin{\footnote{Email: beskin@lpi.ru}}, A.A.~Zheltoukhov\\

\vspace{1cm}

{\it Lebedev Physical Institute, Russian Academy of Sciences,\\ 
Leninskii pr. 53, Moscow, 119991 Russia}\\

\vspace{.4cm}

Received October 10, 2012 \\

\vspace{1cm}

{\small \textit{Pis'ma v Astronomicheskii Zhurnal} \textbf{39}, 
243-248 (2013)
[in Russian]\\
English translation: \textit{Astronomy Letters}, 
\textbf{39}, 215-220 (2013)
\\
 Translated by  V. Astakhov
}

\end{center}

\vspace{1cm}

{\small
{\bf Abstract} -- Using the Grad-Shafranov equation, we consider a new analytical model of the 
black hole magnetosphere based on the assumption that the magnetic field is radial near the 
horizon and uniform (cylindrical) in the jet region. Within this model, we have managed to show 
that the angular velocity of particles $\Omega_{\rm F}$  near the rotation axis of the 
black hole can be smaller than $\Omega_H/2$. This result is consistent with the latest 
numerical simulations.
}
\noindent

\vspace{3cm}
\noindent
DOI: 10.1134/S1063773713040014

\newpage

\begin{center}

{\Large 1. INTRODUCTION}

\end{center}

The main model responsible for the energy release
in active galactic nuclei (AGNs) is presently known to
be the electrodynamic model dating back to the paper
by Blandford and Znajek (1977). Within this model,
the energy losses of a rotating black hole are related
to the flux of electromagnetic energy flowing along the
magnetic field lines from the black hole surface in the
direction of the jets. The question about the magnetic
field structure in the vicinity of a black hole (which
should be generated in an accretion disk) still remains
an open one. This question becomes particularly
topical in connection with the latest observations of
the inner jet regions (see, e.g., Doeleman et al. 2012)
and with the successful launch of the Spectrum-R
(Radioastron) Space Observatory. The latter allows
spatial scales comparable to the size of the central
black hole to be resolved (Kardashev 2009).

By now, a wide variety of magnetic field geometries
near a black hole have been considered in the force-free
approximation within an analytical approach.
These include a split monopole field near the horizon
and far from the black hole (Blandford and Znajek
1977), a parabolic field near the horizon and far
from the black hole (Blandford and Znajek 1977;
Ghosh and Abramowicz 1997), and a uniform magnetic
field near the horizon and a split monopole
field at large distances (Beskin et al. 1992). In all
cases, the angular velocity of the plasma $\Omega_{\rm F}(0)$ 
(which is known to depend uniquely on the magnetic field
geometry) near the rotation axis has always been
exactly half the angular velocity of the black hole
$\Omega_{\rm H}$ (see Fig. 1). However, the latest numerical
simulations show that the condition $\Omega_{\rm F}(0) = 0.5 \, \Omega_{\rm H}$  
can be violated. In particular, McKinney et al. (2012)
argue that their angular velocity profile $\Omega_{\rm F}(\theta)$ near
the black hole horizon more closely corresponds to a
parabolic field, for which actually drops to $0.3 \, \Omega_{\rm H}$
at $\theta = \pi/2$. In this case, however, some of the
magnetic field lines should be connected not with the
black hole horizon but with the accretion disk near
the ergosphere (Punsly 2001).

\begin{figure}
 \centering{\tiny .}\epsfig{figure=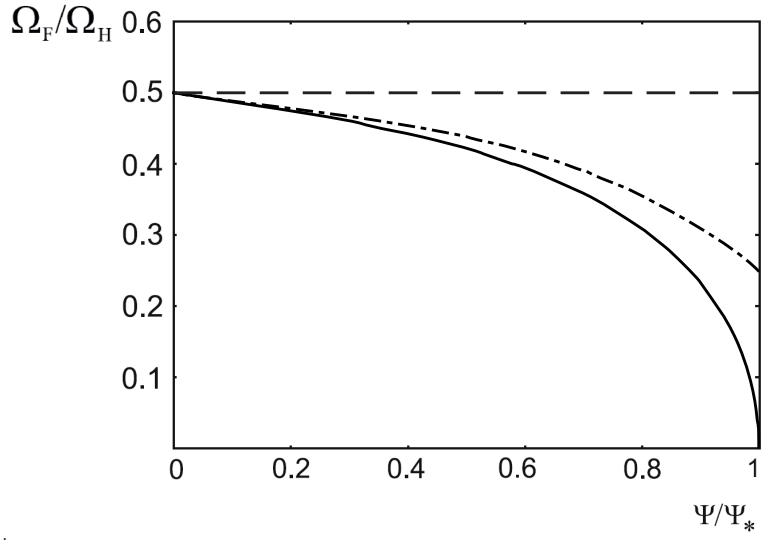,width=0.6 \linewidth}
\caption{ $\Omega_{\rm F}/\Omega_H$ versus $\Psi/\Psi_{\star}$ for a split monopole 
field near the horizon and far from the black hole (dashed line), a parabolic 
field near the horizon and far from the black hole (dash-dotted line), and an uniform 
magnetic field near the horizon and a split monopole field at large distances 
(solid line). Here $\Psi_{\star}$ is the total magnetic flux through the 
horizon.}
\end{figure}

In this paper, we study an analytical model of
the black hole magnetosphere based on a previously
unconsidered magnetic field geometry: a radial field
near the horizon and a vertical field far from the black
hole. In the second section, we give a brief overview
of the Grad-Shafranov equation and other models of
the black hole magnetosphere based on it. In the third
section, we consider the model itself and compare it
with the results of numerical simulations of the black
hole magnetosphere (McKinney et al. 2012). We
show that the derived angular velocity profile can be
easily explained in terms of this model.
 
\vspace{.4cm}

\begin{center}
{\Large 2. THE GRAD-SHAFRANOV EQUATION}
\end{center}

The Grad-Shafranov equation describes axisymmetric
stationary flows in terms of ideal magnetohydrodynamics
(MHD). This approximation is based
on the assumption about a good conductivity of the
plasma filling the magnetosphere of a compact astrophysical
object. In the vicinity of a rotating black hole
(whose metric is also axisymmetric and stationary),
this is provided by an efficient production of electron-
positron pairs (Blandford and Znajek 1977). This
approach is convenient in that quite a few integrals
of motion, i.e., quantities conserved along the particle
trajectory, exist in the case of stationary ideal MHD.
This allows the MHD equations to be reduced to one
second-order equation for the magnetic flux function
$\Psi(r,\theta)$, defining the magnetic field:
\begin{equation}
\label{eq1}
{\bf B} =\frac{\nabla \Psi \times {\bf e}_\varphi}{2\pi\varpi}  
- \frac{2I}{c \varpi}{\bf e}_\varphi.
\end{equation}
Here, $\varpi = \sqrt{g_{\varphi\varphi}}$ is the distance to the 
rotation axis. For this choice of designations, the function  $\Psi(r,\theta)$ 
coincides with the magnetic flux passing through a
circle  $r, \theta, 0 <\varphi < 2\pi$
while the function  $I(r,\theta)$ is
the total current flowing through the same circle.

In addition, the following important properties
hold. \\
(1) The equation $\nabla \cdot {\bf B}=0$ holds automatically.
As a result, the three magnetic field components are
defined by two scalar functions,  $\Psi(r,\theta)$ and $I(r,\theta)$. \\
(2) Since the equation ${\bf B} \cdot \nabla \Psi=0$ holds automatically,
the $\Psi(r,\theta)$  = const lines specify the shape
of the magnetic surfaces.

Next, using the freezing-in condition  ${\bf E} + {\bf v} \times {\bf B}/c = 0$
and the assumption about axisymmetry, we
can determine the electric field as follows (for more
details, see Beskin 2010):
\begin{equation}
\label{eq3}
{\bf E} =- \frac{(\Omega_{\rm F} - \omega)}{2\pi c}\nabla\Psi,
\end{equation}
where $\omega$ is the Lense-Thirring angular velocity. As a
result, the Maxwell equation  $\nabla \times {\bf E}=0$ leads to the
relation $\nabla\Omega_{\rm F} \times \nabla\Psi=0$, whence it follows that
\begin{equation}
\label{eq3}
\Omega_{\rm F} = \Omega_{\rm F}(\Psi).
\end{equation}

The function $\Omega_{\rm F}$ introduced in this way means the
angular velocity of the particles moving in a plasma filled
magnetosphere, while condition (3) is Ferraro 
isorotation law, according to which the angular velocity
of the particles on axisymmetric magnetic surfaces
must be constant (Ferraro 1937). Similarly, from the
Maxwell equations we can deduce that $\nabla I \times \nabla\Psi=0$
and, consequently,
\begin{equation}
\label{eq4}
I=I(\Psi).
\end{equation}
This means that the total electric current within a
magnetic flux tube is also conserved.

It is important to emphasize that in contrast to
the nonrelativistic problem, there is a second family
of singular surfaces associated with the accreting
matter in the black hole magnetosphere. As a result,
the additional critical condition allows an additional
relation between the current $I(\Psi)$ and angular velocity
$\Omega_{\rm F}(\Psi)$ to be determined. In the force-free
appro\-xi\-ma\-tion, this relation can be written as (Thorne
and MacDonald 1982)
\begin{equation}
\label{eq9}
4\pi I(\Psi)=\left[\Omega_H-\Omega_{\rm F}(\Psi)\right]\sin\theta
\frac{r_{\rm g}^2+a^2}{r_{\rm g}^2+a^2\cos^2\theta}
\left(\frac{{\rm d}\Psi}{{\rm d}\theta}\right),
\end{equation}
where $r_{\rm g}$ is the black hole radius and  $a$ is the rotation
parameter. Recall that the true meaning of Eq. (5) is
the critical condition on the inner fast magnetosonic
surface that coincides with the black hole horizon in
the force-free approximation (Beskin 2010). As a
result, this condition allows not only the longitudinal
current $I(\Psi)$ but also the angular velocity  $\Omega_{\rm F}(\Psi)$ to be
determined.

\vspace{.4cm}

\begin{center}
{\Large 3. THE PLASMA ANGULAR VELOCITY PROFILE}
\end{center}

As has already been said, several analytical models
of the black hole magnetosphere were proposed in
the literature. The first of them was constructed by
Blandford and Znajek (1977). They considered a
slowly rotating black hole for which a nonrotating
black hole with a split monopole field was chosen as
the zeroth approximation. Such a geometry can be
easily realized in the presence of a thin accretion disk.
In this case, the flux function  $\Psi=\Psi_0 (1-\cos\theta)$
for \mbox{$\theta < \pi/2$} and $\Psi=\Psi_0 (1+\cos\theta)$ for
$\theta > \pi/2$ will be an
exact solution of the Grad-Shafranov equation for a
nonrotating black hole. The same authors considered
a model magnetosphere with a parabolic magnetic
field in the vicinity of a slowly rotating black hole. The
shape of the field lines for $\theta < \pi/2$  at large distances
is described by the flux function  
\mbox{$\Psi=\Psi_0 (r/r_{\rm g})(1-\cos\theta)$.}
Since $\Psi(r,\pi)\neq $ const for it, this implies the presence
of sources or sinks in the volume (and not only in
the gravitating center or at infinity). Such sources
can also be realized in an accretion disk. Finally,
Beskin et al. (1992) investigated the case where the
black hole is in the center of a well-conducting disk
bounded by an inner radius $b$. In this case, the magnetic
field was almost uniform near the black hole and
still remained a split monopole one at large distance
$(r \gg b)$. As can be seen from Fig. 1, the angular
velocity $\Omega_{\rm F}(\Psi)$  near the rotation axis 
is $\Omega_{\rm H}/2$ in all these cases.

\begin{figure}
\centering{\tiny .}\epsfig{figure=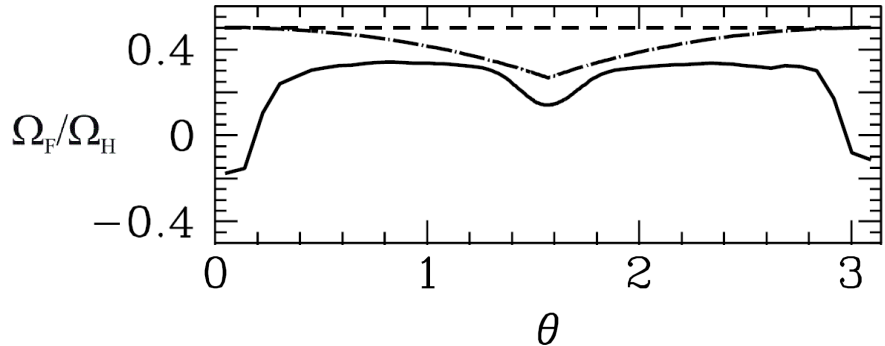,width=0.7 \linewidth}
\caption{The plot of $\Omega_{\rm F}/\Omega_H$ 
on the horizon versus polar angle derived during the numerical
simulations of the black hole
magnetosphere (McKinney et al. 2012). The dashed and dash-dotted 
lines correspond to monopole and parabolic fields,
respectively.
}
\end{figure}

On the other hand, as is shown in Fig. 2, in their
recent paper devoted to numerical simulations of the
black hole magnetosphere, McKinney et al. (2012)
obtained a profile of the angular velocity $\Omega_{\rm F}$ 
that not only differed from  $\Omega_{\rm H}/2$ near the 
axis but also even became negative here. The authors conclude 
that the derived profile is nevertheless closer to the parabolic
solution, especially since the outer magnetic surfaces
actually have such a shape. However, a significant
fraction of the magnetic field lines in the parabolic solution
must pass through the accretion disk. Consequently,
not the rotating black hole but the equatorial
region of the ergosphere will be an energy source for
the corresponding magnetic surfaces (such a model
was developed, for example, in Punsly's works; see
Punsly 2001).

Below, we will show that the results obtained by
McKinney et al. (2012) are in best agreement with
the previously unconsidered model of a black hole
magnetosphere with a (split) monopole magnetic field
near the black hole horizon and a cylindrical field far
from the black hole. In this model, in which the bulk of
the magnetic flux now passes through the black hole
horizon, not only a collimation of the magnetic surfaces
(it will be connected with the fairly high density
of the ambient medium) but also angular velocities
smaller than $\Omega_{\rm H}/2$ near the axis can be obtained.

In our model, we will use the assumption that the
flow is cylindrical near the rotation axis far from the
black hole, as is obtained in numerical simulations,
so that all quantities depend only on the cylindrical
radius $\varpi$. In this case, the Grad-Shafranov equation
is one-dimensional and can be integrated. In the
force-free approximation, the solution of the Grad-Shafranov 
equation takes the form (see, e.g., Istomin and Pariev 1994)
\begin{equation}
\label{eq6}
\Omega_{\rm F}^2(\Psi)\varpi^4 B_z^2 c^{-2}=\varpi^2 B_\varphi^2 + 
\int\limits_0^\varpi x^2 \frac{{\rm d}}{{\rm d}x} \left(B_z\right)^2{\rm d}x.
\end{equation}
Substituting the expression of the toroidal field in
terms of the total current $B_\varphi=-{2I}/{\varpi c}$, 
we can rewrite Eq. (6) as
\begin{equation}
\label{eq7}
\Omega_{\rm F}^2(\Psi)A_1^2(\Psi)=4I^2(\Psi) + A_2(\Psi), 
\end{equation}
where the following notation is used:
\begin{eqnarray}
\label{eq8}
A_1(\Psi) & = & \varpi^2 B_z; \\
A_2(\Psi) & = & c^2\int\limits_0^\varpi x^2 \frac{{\rm d}}{{\rm d}x} \left(B_z\right)^2{\rm d}x.
\end{eqnarray}

As regards the "boundary condition on the horizon"
(5), it can be rewritten as
\begin{eqnarray}
\label{eq10}
2I(\Psi) & = & \left[\Omega_H-\Omega_{\rm F}(\Psi)\right]A_3(\Psi), 
\end{eqnarray} 
where
\begin{eqnarray}
\label{eq10'}
A_3 & = & \frac{1}{2\pi}\sin\theta\frac{r_{\rm g}^2+a^2}{r_{\rm g}^2+a^2\cos^2\theta}
\left(\frac{{\rm d}\Psi}{{\rm d}\theta}\right).
\end{eqnarray} 

Substituting the latter expression for $2I(\Psi)$ into
Eq. (7) yields a quadratic equation for $\Omega_{\rm F}(\Psi)$:
\begin{equation}
\label{eq11}
\Omega_{\rm F}^2 (A_1^2-A_3^2)+2\Omega_{\rm F}\Omega_H A_3^2 - \Omega_H^2 A_3^2 - A_2=0.
\end{equation}
Hence, the general expression for the angular velocity
can be written as
\begin{equation}
\label{eq13}
\Omega_{\rm F}=\Omega_{\rm H}\left[\frac{A_3}{A_3+A_1} +
\frac{A_2}
{\Omega_{\rm H}^2 A_1 A_3 \left(1 + \sqrt{1-\frac{A_2(A_3^2-A_1^2)}{\Omega_{\rm H}^2 A_1^2 A_3^2}}\right)
}\right].
\end{equation}
This form stems from the fact that the relation $A_1 =A_3$ holds for the previously 
considered solutions on
the rotation axis. Therefore, we tried to avoid the
quantities $(A_1 - A_3)$ in the denominators of the corresponding
expressions.
Let us first consider the case where the magnetic
field is vertical and uniform far from the black hole
($\Psi = \pi \varpi^2 B_{0}$ at $r\gg r_{\rm g}$) and exactly radial on the
horizon ($\Psi = \Psi_{*}(1-\cos\theta)$ at $r = r_{\rm g}$). Substituting
the corresponding flux functions into Eqs. (8), (9),
and (11), we will then obtain
\begin{eqnarray}
\label{eq14}
A_1(\Psi) & = & \varpi^2 \frac{1}{2\pi\varpi}\frac{{\rm d}\Psi}{{\rm d}\varpi}=\frac{\Psi}{\pi},
\nonumber \\
A_2(\Psi) & = & c^2\int\limits_0^\varpi x^2 \frac{{\rm d}}{{\rm d}x} \left(B_0\right)^2{\rm d}x =0,
\nonumber \\
A_3(\Psi) & = & \frac{\Psi}{\pi}\cdot\frac{r_{\rm g}^2+a^2}{r_{\rm g}^2+a^2(1-\Psi/\Psi_{*})^2}.
\end{eqnarray}
Consequently, $\Omega_{\rm F} = \Omega_{\rm H}/2$ on the rotation axis.
On the other hand, both analytical (Beskin and
Nokhrina 2009) and numerical (Komissarov et al.
2006; Tchekhovskoy et al. 2009; Porth et al. 2011)
calculations show that a central core, which, as will
be shown below, can change significantly the situation,
can exist in the jet. Consider the case where the
magnetic field is still exactly radial on the horizon and
vertical far from the black hole, but now a denser core
of radius $r_{\rm core}$ exists near the rotation axis. As has
been shown, such a core must actually be formed at a
sufficiently low pressure of the ambient medium, with
(see, e.g., Beskin and Nokhrina 2009)
\begin{equation}
\label{eq15}
r_{\rm core} =  k \frac{c}{\Omega_{\rm F}(0)}.
\end{equation}
Here, $k \approx \gamma_{\rm in}$, where $\gamma_{\rm in}$ is the characteristic Lorentz
factor of the particles flowing along the jet axis. At
distances $\varpi \leq r_{\rm core}$ from the rotation axis, we can
then write
\begin{equation}
\label{eq16}
B_z=B_0 - B_0 \frac{\varpi^2}{r^2_{\rm core}}.
\end{equation}
The following flux function corresponds to this field:
\begin{equation}
\label{eq17}
\Psi= \pi \varpi^2 B_{0} - \frac{1}{2}\pi B_{0} \frac{\varpi^4}{r^2_{\rm core}}.
\end{equation}
As a result, we still have $A_1(\Psi) \approx A_3(\Psi) \approx \Psi/\pi$
near the rotation axis in the first order in $\Psi$. Now, however,
$A_2$ will be nonzero,
\begin{equation}
\label{eq18}
A_2\approx -\frac{\Psi^2}{\pi^2 r_{\rm core}^2}.
\end{equation}
In the upshot, substituting the expressions for $A_1, A_2$
and $A_3$ into the general formula (13), we have
\begin{equation}
\label{eq19}
\Omega_{\rm F}(0) \approx \frac{\Omega_{\rm H}}{2}\left(1 -
\frac{c^2}{\Omega_{\rm H}^2 r_{\rm core}^2
}\right).
\end{equation}
If, however, we express here $r_{\rm core}$ using Eq. (15), then
\begin{equation}
\label{eq19next}
\Omega_{\rm F}(0) \approx \frac{\Omega_{\rm H}}{1 + \sqrt{1+1/k^2}}.
\end{equation}
As we see, in the presence of a dense core, the angular
velocity on the jet axis $\Omega_{\rm F}(0)$ can be smaller than $\Omega_{\rm H}/2$.
In particular, for a mildly relativistic flow, $k = 1$, we obtain 
\mbox{$\Omega_{\rm F}(0) = 0.41 \, \Omega_{\rm H}$}.

Let us now use our model to analyze the results
of the numerical simulations performed by McKinney
et al. (2012), in which a dense core also takes place.
According to these results, the magnetic field near
the black hole horizon may be considered radial with
a good accuracy. From the plot of $B_r(r_{\rm H},\theta)$, we
can then derive the plot of the magnetic flux function
$\Psi(r_{\rm H},\theta)$ and subsequently the plot of $A_3(\Psi)$. Since
the flow near the axis at large distances may be considered
cylindrical with a good accuracy, we may set
$B_z\approx B_r$ in this region and use the plot of the radial
magnetic field at $r=30 r_g$. However, the magnetic
field directly on the axis is overestimated due to the
peculiarities of the numerical method.
Let us now consider a model magnetic field
\begin{equation}
\label{eq21}
B_z=\frac{B_0}{1+\varpi^2/r^2_{\rm core}}+B_1,
\end{equation}
where $B_0, B_1$ and $r_{\rm core}$  are the parameters of the
problem. Let us choose them so that, first, the plot
of the function $B_z(\theta)$ at small $\theta$ is close to the plot of
$B_r(30 r_g,\theta)$ from McKinney et al. (2012), and, second,
the total magnetic flux corresponding to this $B_z$
should coincide with the total magnetic flux on the
horizon. The latter condition is based on the property
of magnetic flux conservation and the fact that
much of the magnetic flux emerging from the black
hole horizon is subsequently concentrated inside the
jet, i.e., near the axis. The following flux function
corresponds to the model magnetic field (21):
\begin{equation}
\label{eq22}
\Psi=\pi r^2_{\rm core}B_0 \ln(1+\varpi^2/r^2_{\rm core})+\pi B_1 \varpi^2.
\end{equation}
Such a flux function does not allow the inverse dependence
$\varpi(\Psi)$ and, consequently, the dependences
$A_1(\Psi)$ and $A_2(\Psi)$ to be derived analytically. However,
these dependences can be derived numerically and the
angular velocity profile $\Omega_{\rm F}(\Psi(\varpi)$ can be found using
Eq. (13).

\begin{figure}
 \centering{\tiny .}\epsfig{figure=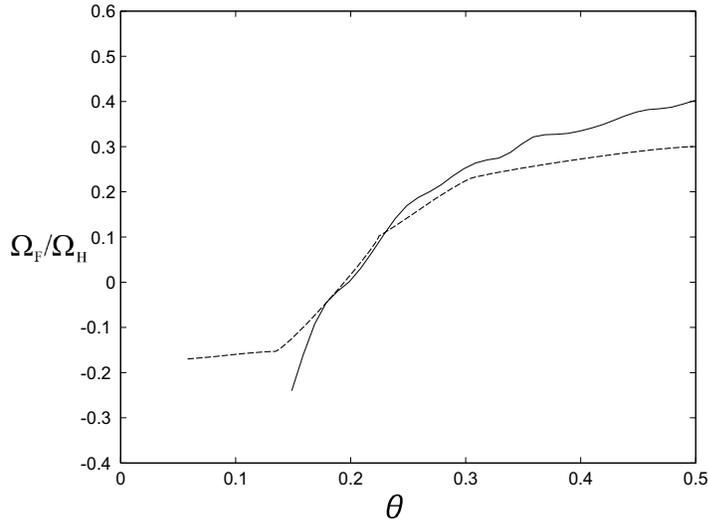,width=0.6 \linewidth}
\caption{
The plot of $\Omega_{\rm F}/\Omega_H$ on the horizon versus polar angle $\theta$
derived from the analytical model described here (solid line) and
the plot from McKinney et al. (2012) (dashed line).
}
\end{figure}

Fig. 3 presents a plot where $\Omega_{\rm F}/\Omega_{\rm H}$ on the black
hole horizon is along the vertical axis and the polar
angle $\theta$ is along the horizontal axis. The range of
angles was chosen from the following considerations.
At very small $\theta$, the magnetic field obtained in the
numerical simulations diverges, most likely due to the
peculiarities of the numerical method. On the other
hand, at large $\theta$, the assumption that the magnetic
field is vertical will break down. As we see, the
proposed model is in excellent agreement with the
numerical simulations. Note that $\Omega_{\rm F}$ is negative near
the axis, which is also in agreement with the work by
McKinney et al. (2012).

\vspace{.4cm}

\begin{center}
{\Large 4. CONCLUSIONS}
\end{center}

We investigated a new analytical model of the
black hole magnetosphere based on a previously unconsidered
geometry of magnetic surfaces: a radial
magnetic field near the horizon and a vertical field far
from the black hole. We showed that in the presence
of a dense core near the jet axis, there is excellent
agreement of this model with the numerical simulations.
And this is despite the fact that the analytical
calculations were performed within the simplest
force-free approximation and under the assumption
that the flow was axisymmetric and stationary, while
McKinney et al. (2012) carried out their 3D numerical
simulations in the full MHD version by taking into
account the fact that the flows under consideration
were nonstationary.

We emphasize that the negative values of the angular
velocity $\Omega_{\rm F}$ are most likely associated with the
difficulties of the numerical procedure near the rotation
axis. Therefore, actually one should not expect
the appearance of a region with counter-rotation near
the jet axis. For us it was important here to only
show that given the magnetic field structure near the
black hole horizon and in the jet region, the angular
velocity profile obtained in a self-consistent way could
be reproduced using a simple analytical model.

Good agreement between the theory and numerical
simulations once again shows that axisymmetric
stationary flows, for which quite a few analytical results
have been obtained in the last three decades, remain
a good basis for analyzing the processes occurring
in real astrophysical sources. One of such properties
is that despite the turbulent nature of the flow in
the region above the accretion disk, the flow near the
rotation axis remains fairly regular. Therefore, there is
hope that the previously formulated simple analytical
asymptotics (and, in particular, the assertion that the
magnetic field structure near the horizon should be
nearly radial) will also be needed in the future.

Finally, note that a parabolic field such that a
significant fraction of the magnetic field lines cross
the equator within the ergosphere would require the
existence of an energy source directly in the accretion
disk. In the steady-state problem, such a situation
is unlikely to be possible. In our view, the fact that
the 3D simulations discussed above lead to a quasiparabolic
structure of magnetic surfaces is related to
a fairly high external pressure. As a result, the jet
radius exceeds the black hole horizon radius only by
several times. In fact, however, as can be clearly seen
from the structure of the innermost magnetospheric
regions in McKinney et al. (2012), only a very small
fraction of the magnetic field lines pass through the
equator.

\vspace{.4cm}

\begin{center}
{\Large ACKNOWLEDGMENTS}
\end{center}

We are grateful to A. Tchekhovskoy for the provided
numerical simulations, A.A. Philippov for a
helpful discussion, and the Ministry of Education
and Science of the Russian Federation for financial
support (Contract No. 8525).

\vspace{.4cm}

\begin{center}
{\Large REFERENCES}
\end{center}

\noindent
V. S. Beskin, Phys. Usp. 53, 1199 (2010). \\
V. S. Beskin, Ya. N. Istomin, and V. I. Pariev, Sov.
Astron. 36, 642 (1992).                        \\
V. S. Beskin and E. E. Nokhrina, Mon. Not. R. Astron.
Soc. 397, 1486 (2009).                           \\
R. D. Blandford and R. L. Znajek, Mon. Not. R. Astron.
Soc. 179, 433 (1977).                              \\
S. S. Doeleman, V. L. Fish, D. E. Schenck, et al.,
Science 338, 355 (2012).                             \\
V. C. A. Ferraro, Mon. Not. R. Astron. Soc. 97, 458
(1937).                                                \\
P. Ghosh and M. A. Abramowicz, Mon. Not. R. Astron.
Soc. 292, 887 (1997).                                    \\
Ya. N. Istomin and V. I. Pariev, Mon. Not. R. Astron.
Soc. 267, 629 (1994).                                      \\
N. S. Kardashev, Phys. Usp. 52, 1127 (2009).          \\
S. Komissarov, M. Barkov, N. Vlahakis, and               
A. K\"onigl, Mon. Not. R. Astron. Soc. 380, 51 (2006).        \\
J. C. McKinney, A. Tchekhovskoy, and R. D. Blanford,
Mon. Not. R. Astron. Soc. 423, 2083 (2012).                    \\
O. Porth, Ch. Fendt, Z.Meliani, and B. Vaidya, Astrophys.
J. 737, 42 (2011).                                                \\
B. Punsly, Black Hole Gravitohydromagnetics                  
(Springer, Berlin, 2001).                                        \\
A. Tchekhovskoy, J. McKinney, and R. Narayan, Astrophys.
J. 699, 1789 (2009).                                               \\
K. S. Thorne and D. MacDonald, Mon. Not. R. Astron.
Soc. 198, 339 (1982).                                               

\end{document}